\documentclass[aps,pre,preprint,showpacs,byrevtex]{revtex4}
\usepackage{bm}
\begin{document}

\title{A second constant of the motion for two-dimensional positronium in a magnetic field}

\author{Gerardo Mu\~{n}oz}
\affiliation{Department of Physics\\
California State University, Fresno\\
Fresno, CA 93740-0037\\
\\}


\begin{abstract}
Recent numerical work indicates that the classical motion of positronium in a constant magnetic field does not exhibit chaotic behavior if the
system is confined to two dimensions. One would therefore expect this system to possess a second constant of the motion in addition to the total
energy. In this paper we construct a generalization of the Laplace-Runge-Lenz vector and show that a component of this vector is a constant of the
motion.
\end{abstract}
\pacs{36.10.Dr, 45.50.Dd}
\maketitle

Anderson, Murawski, and Schmidt\cite{AMS} have recently performed a numerical study of the classical motion of positronium embedded in a constant
magnetic field. They find that the three-dimensional system is chaotic, suggesting the possibility of a chaos-assisted tunneling reduction of the
lifetime of positronium which Ackermann, Shertzer, and Schmelcher\cite{Ackermann, Shertzer} had previously predicted to be on the order of years.
When the system is confined to two dimensions, however, no sign of chaos emerges from their computations. Indeed, they were able to show that the
largest Lyapunov exponent is zero. These results suggest the existence of an additional constant of the motion (CM). We show here that such a CM
may be obtained from a component of a generalized Laplace-Runge-Lenz vector\cite{Laplace, Runge, Lenz}.

For completeness, we review the basic equations in the notation of Ref.\cite{AMS}. The system consists of two particles of equal mass $m$ and
charges $+e$ and $-e$ moving nonrelativistically in a constant magnetic field ${\bf B}$:
\begin{equation}
m\: {\bf \ddot r}_1 \; = \; e\:{\bf \dot r}_1 \times {\bf B} \:-\: {e^2 \over 4 \pi \epsilon_0}{{\bf r}_1 - {\bf r}_2 \over |{\bf r}_1 - {\bf
r}_2|^3}
\end{equation}
\begin{equation}
m\: {\bf \ddot r}_2 \; = \; -e\:{\bf \dot r}_2 \times {\bf B} \:+\: {e^2 \over 4 \pi \epsilon_0}{{\bf r}_1 - {\bf r}_2 \over |{\bf r}_1 - {\bf
r}_2|^3}
\end{equation}

One may easily integrate the sum of Eqs. (1) and (2) to find the conserved quantity (called the pseudomomentum in Refs.\cite{Ackermann, Shertzer})
\begin{equation}
m\: {\bf \dot R} - e\:{\bf r} \times {\bf B} \; = \; {\bm \alpha}
\end{equation}
where
\begin{eqnarray}
{\bf R} \; = \; {\bf r}_1 + {\bf r}_2 \\
{\bf r} \; = \; {\bf r}_1 - {\bf r}_2
\end{eqnarray}

The difference of Eqs. (1) and (2) leads to
\begin{equation}
m\: {\bf \ddot r} \; = \; e\:{\bf \dot R} \times {\bf B} \:-\: {e^2 \over 2 \pi \epsilon_0}{{\bf r} \over r^3}
\end{equation}
Substitution of ${\bf \dot R}$ from Eq. (3) yields an equation for ${\bf r}$,
\begin{equation}
m\: {\bf \ddot r} \; = \; {e \over m}\:( e\:{\bf r} \times {\bf B} +  {\bm \alpha}) \times {\bf B} \:-\: {e^2 \over 2 \pi \epsilon_0}{{\bf r}
\over r^3}
\end{equation}
Choosing the coordinate system such that ${\bf B} = B {\bf e}_3$ and rescaling $t$ and ${\bf r}$ according to $(eB/m)t \rightarrow t$,
$(2 \pi \epsilon_0 B^2/m)^{1/3}{\bf r} \rightarrow {\bf r}$ as in Ref.\cite{AMS} one finds
\begin{equation}
{\bf \ddot r} \; = \; ({\bf r} \times {\bf e}_3 )\times {\bf e}_3 - {{\bf r} \over r^3} + {\bm \beta}
\end{equation}
where ${\bm \beta} = (2 \pi \epsilon_0 /m e^3 B)^{1/3} {\bm \alpha} \times {\bf e}_3$.

In the two-dimensional case, $({\bf r} \times {\bf e}_3 )\times {\bf e}_3 = -{\bf r}$, and the above simplifies to
\begin{equation}
{\bf \ddot r} \; = \; -{\bf r} - {{\bf r} \over r^3} + {\bm \beta}
\end{equation}
The presence of the harmonic term on the right-hand side makes ionization impossible for two-dimensional positronium.

Eq. (9) implies that the dimensionless angular momentum ${\bf L} = {\bf r} \times {\bf \dot r}$ is not conserved. Indeed, taking the cross
product of ${\bf r}$ with Eq. (9) we have
\begin{equation}
{\bf r} \times {\bf \ddot r} \; = \; {\bf r} \times {\bm \beta}
\end{equation}
or
\begin{equation}
{\bf  \dot  L} \; = \; {\bf r} \times {\bm \beta}
\end{equation}

A fairly obvious constant of the motion is the dimensionless energy
\begin{equation}
E = \frac12 \, v^2 + \frac12 \, r^2 - \frac1r - {\bm \beta} \cdot{\bf r}.
\end{equation}
A second nontrivial CM may be obtained by taking the cross product of Eq. (11) with ${\bf \dot r}$, and then computing the scalar product of the
resulting equation with ${\bm \beta}$. After taking this cross product, adding and subtracting ${\bf L} \times {\bf \ddot r}$ on the left-hand
side, and using Eq. (9) we get
\begin{equation}
{d \over dt} ({\bf L} \times {\bf \dot r}) + {\bf L} \times ({\bf r} + {{\bf r} \over r^3} - {\bm \beta}) = ({\bf r} \times {\bm
\beta})\times {\bf \dot r}
\end{equation}
But ${\bf L} \times {\bf r}/ r^3 = d {\bf \hat r}/dt$, where ${\bf \hat r} = {\bf r}/r$ is the unit vector in the ${\bf r}/r$ direction. Therefore
\begin{equation}
{d{\bf A} \over dt} + {\bf L} \times ({\bf r} - {\bm \beta}) = ({\bf r} \times {\bm \beta})\times {\bf \dot r}
\end{equation}
with ${\bf A} = {\bf \hat r} + {\bf L} \times {\bf \dot r}$ the dimensionless Laplace-Runge-Lenz vector. We now take the dot product of Eq. (14)
with ${\bm\beta}$:
\begin{eqnarray}
{d \over dt}({\bf A} \cdot {\bm \beta}) &=& - ({\bf L} \times {\bf r}) \cdot {\bm \beta} \:+\: [({\bf r} \times {\bm \beta})\times {\bf \dot
r}]\cdot {\bm \beta} \nonumber \\
&=& - {\bf L} \cdot ({\bf r} \times {\bm \beta})\:+\: ({\bf r} \times {\bm \beta}) \cdot ({\bf \dot r} \times {\bm \beta}) \nonumber\\
&=& - {\bf L} \cdot{\bf  \dot  L} + {\bf  \dot  L} \cdot{\bf  \ddot  L} \nonumber
\end{eqnarray}
Thus
\begin{equation}
{d \over dt} \left({\bf A} \cdot {\bm \beta} \:+\: \frac12 \: {\bf L}^2 - \frac12 \: {\bf  \dot  L}^2 \right) = 0
\end{equation}
or
\begin{equation}
{\bf A} \cdot {\bm \beta} \:+\: \frac12 \: ({\bf r} \times {\bf \dot r})^2 - \frac12 \: ({\bf r} \times {\bm \beta})^2 = C_{\beta}
\end{equation}
with $C_{\beta}$ a constant. It is easy to show that the above is equivalent to the statement that the scalar product of ${\bm \beta}$ with
the vector\cite{Redmond}
\begin{equation}
{\bf C} \:=\: {\bf A} \:-\: {1 \over 2 \beta^2} \: {\bf L} \times({\bf L} \times {\bm \beta})\:+\: \frac12 \: {\bf r} \times({\bf r}
\times {\bm \beta}) \:=\: {\bf A} \:+\: {L^2 \over 2 \beta^2} \: {\bm \beta}\:+\: \frac12 \: {\bf r} \times({\bf r}
\times {\bm \beta})
\end{equation}
is conserved. Note that the vector itself is not conserved, since
\begin{equation}
{d{\bf C} \over dt} \:=\: \left( \frac32 \: \beta - {\bf r} \cdot {\bm {\hat\beta}} \right){\bf L} \times {\bm {\hat\beta}}
\end{equation}

The existence of the new CM $C_{\beta} = {\bf C} \cdot {\bm \beta}$ depends crucially on the equality of the masses of
the particles involved; it no longer obtains if
$m_1 \neq m_2$. Furthermore, it is an exact conservation law for the classical dynamics of two-dimensional positronium only if the experimental
conditions warrant modeling the real system by Eqs. (1), (2), i.e., only if the conditions are such that spin, radiative, and relativistic effects
are irrelevant. On the other hand, such effects are expected to be small for the delocalized states of interest in Refs.\cite{Ackermann,
Shertzer}. Hence, even if the improved (by the inclusion of spin, radiative, and relativistic effects) description did display chaotic
behavior, Eq. (16) should still represent an approximate conservation law, implying that the predictions of Refs.\cite{Ackermann,Shertzer} would
not be threatened by chaos-assisted tunneling in the two-dimensional case.

\end{document}